\documentclass[%
 aip,
 amsmath,amssymb,
 reprint,%
]{revtex4-1}

\usepackage{ulem}
\usepackage{tabularx}
\usepackage{multirow}
\usepackage{graphicx}
\usepackage{dcolumn}
\usepackage{bm}
\usepackage{epsfig} 
\usepackage{graphicx}
\usepackage{soul}
\usepackage{color}
\usepackage{booktabs}
\usepackage{amsmath}
\usepackage{algorithm}
\usepackage{algorithmicx}
\usepackage[noend]{algpseudocode}	
\usepackage{array}
\usepackage{subfigure}
\usepackage{framed} 
\usepackage[title]{appendix}
\usepackage{ifthen}
\usepackage{nomencl} 
\usepackage{setspace}
\makenomenclature
\usepackage{enumitem}
\usepackage{dsfont}

\DeclareMathOperator{\erf}{erf}
\renewcommand{\thefootnote}{\roman{footnote}}
\newcommand{\cyr}[1]{{\color{black} #1}}

\makeatother

\begin{document}

\preprint{AIP/123-QED}

\title[Complex-Valued Time Series Based Solar Irradiance Forecast]{Complex-Valued Time Series Based Solar Irradiance Forecast}
\author{Cyril Voyant} %
\email{voyant$\_$c@univ-corse.fr}
\affiliation{University of Corsica, SPE Laboratory-Georges Peri'centre - UMR6134, Ajaccio (France)}%

\author{Philippe Lauret}
\affiliation{University of Reunion, PIMENT Laboratory, Saint-Pierre (France)}
\author{Gilles Notton}
\affiliation{University of Corsica, SPE Laboratory-Georges Peri'centre - UMR6134, Ajaccio (France)}%
\author{Jean-Laurent Duchaud}
\affiliation{University of Corsica, SPE Laboratory-Georges Peri'centre - UMR6134, Ajaccio (France)}%
\author{Luis Garcia-Gutierrez}
\affiliation{University of Lorraine, LMOPS Laboratory , Metz (France)}%
\author{Ghjuvan Antone Faggianelli}
\affiliation{University of Corsica, SPE Laboratory-Georges Peri'centre - UMR6134, Ajaccio (France)}%

\date{\today}

\begin{abstract}

 A new method for short-term probabilistic forecasting of global solar irradiance from complex-valued time series is explored. Measurement defines the real part of the time series while the estimate of the volatility is the imaginary part. A complex autoregressive model (capable to capture quick fluctuations) is then applied with data gathered on Corsica island (France).
 Results show that even if this approach is easy to implement and requires very little resource and data, both deterministic and probabilistic forecasts generated by this model are in agreement with experimental data (root mean square error ranging from 0.196 to 0.325 considering all studied horizons). In addition, it exhibits sometimes a better accuracy  than classical models such as the Gaussian process, bootstrap methodology, or even more sophisticated models such as quantile regression. Many studies and many fields of physics could benefit from this methodology and from the many models that could result from it. 
\end{abstract}

\maketitle


\section{Introduction}
\label{sec:introduction}

Nowadays, it is acknowledged that to limit the impact of the random and variable nature of the solar resource and thus to facilitate its integration, developments are necessary. They concern the energy storage means, the smart grid energy management, and the forecasting methods for both power generation and user’s consumption \cite{notton_intermittent_2018}. 
The topic of this paper falls within the development of a forecasting method for Photovoltaic ($PV$) power generation and concerns nowcasting. Numerous machine learning methods benchmarks have been published in the literature and most of them compare the models in terms of accuracy  \cite{ZHOU2021113960,voyant_machine_2017} with regard to time horizons. These methods capture often the general trend and fail to capture the quick fluctuations, while advanced nonparametric approaches that attempt to do so may be prone to overfitting or too complicated for practical applications (lack of data, acquisition system failures, process execution time, etc.). Many grid managers prefer to use the simplest and the most robust ones, sometimes at the expense of their performance \cite{alamo_advanced_2019}. In agreement with the "No Free Lunch theorem" of \citet{wolpert1977}, which explains that no learning algorithm is the most suitable in all scenarios \cite{cerqueira2019machine}, we propose a new \cyr{data mining based non-parametric} probabilistic method, easy to implement, with good accuracy and based on a new theoretical basis integrating trend but also rapid fluctuations predictions. A univariate methodology based on a complex number generation is applied to predict simultaneously the hourly solar global horizontal irradiance ($GHI$) and an estimate of its volatility from previous ground measurements.

\section{Data}
\label{sec:data}
As detailed by \citet{yang_choice_2020}, an adequate analysis and modeling are essential to issue good forecasts when a time series exhibits seasonal or cyclic behavior as it is the case for $GHI$ with its two seasonal periods (yearly and diurnal cycles). Since 1961 and the first works about stationary processes with a finite second-moment \cite{pagano78} and periodic correlation (or covariance) \cite{gladyshev61}, the scientists know that it is important to pay attention to trends when time series is used. Box and Jenkins' first formalism \cite{box_time_1976} clarified this aspect by proposing a decomposition, especially when seasonality is easily quantifiable. 
Usually, a multiplicative scheme is chosen, and a classical ratio between $GHI$ in clear sky condition (denoted $GHI_{CS}$) and $GHI$ is operated. This parameter \cyr{(considered ``sufficiently'' stationary or at least locally stationary as demonstrated by Yang et al \cite{yang_choice_2020})} results in a normalized quantity (theoretically comprised between $0$ and $1$ as long as the over-irradiance phenomenon is neglected) known as $\kappa(t)$ the clear-sky index  \cite{doi:10.1063/5.0105020},
\begin{equation}
\label{eq:csi}
\kappa(t)=GHI(t)/GHI_{CS}(t)\in[0,1]
\end{equation}

Thus, most solar forecasters build their forecasting models on $\kappa$, rather than on $GHI$ itself. As a part of this study, several rules and explanations must be given to improve the objectivity of conclusions:
\begin{itemize}[label=$\looparrowright$]
\item $GHI$ time series is measured in Ajaccio (Corsica, France, 41.92N-8.74E, 5m above sea level) endowed with a warm Mediterranean climate ($Csa$ Köppen climate classified) and yearly solar irradiation of 1642 kWh.m$^{-2}$, 
\item Models are evaluated during only daytime irradiance values, filtering the checked data (\cyr{less than 1\% are left} according to quality control \cite{app12178529}) on solar zenith angle ($GHI=\emptyset$ if $\theta_Z>85^{\circ}$),
\item $GHI_{CS}$ is computed with the \texttt{Solis} model which proposes an atmospheric scheme based on radiative transfer calculations and the Lambert-Beer relation \cite{ineichen_broadband_2008}.
\end{itemize}

This paper is dedicated to volatility prediction which is used to generate $GHI$ prediction intervals with respect to the prediction horizon from $1h$ to $6h$ with $1h$ time granularity (training during the years 2008-2017 and testing during the year 2018). 
Several methods such as autoregressive conditional heteroskedasticity ($ARCH$) models are devoted to this task (volatility modeling) and were extensively studied in econometrics. However, concerning the $GHI$ prediction and its applications in energy management for PV systems, this kind of method has never been used, probably due to its complexity \cite{dimson_volatility_1990}, the restrictive assumptions \cite{nwogugu_further_2006} or the quality of its results which seemed even so promising \cite{david_probabilistic_2016}. An important conclusion of \citet{dimson_volatility_1990} concerning the $ARCH$ family predictors is another form of the Occam's razor principle and implies ``that for those who are interested in forecasts with reasonable predictive accuracy, the best forecasting models might well be the simplest ones''.

\section{Methodology}
\label{sec:method}
The method exposed in this paper concerns a new formalism for: 
\begin{itemize}[label=$\looparrowright$]
\item The prediction of the conditional volatility using parameters like the \textit{return}  and its \textit{standard deviation} (see definition in Eq.\ref{eq:eq5}),
\item The generation of $GHI$ prediction intervals.
\end{itemize}

\begin{figure}
\includegraphics[scale=0.310]{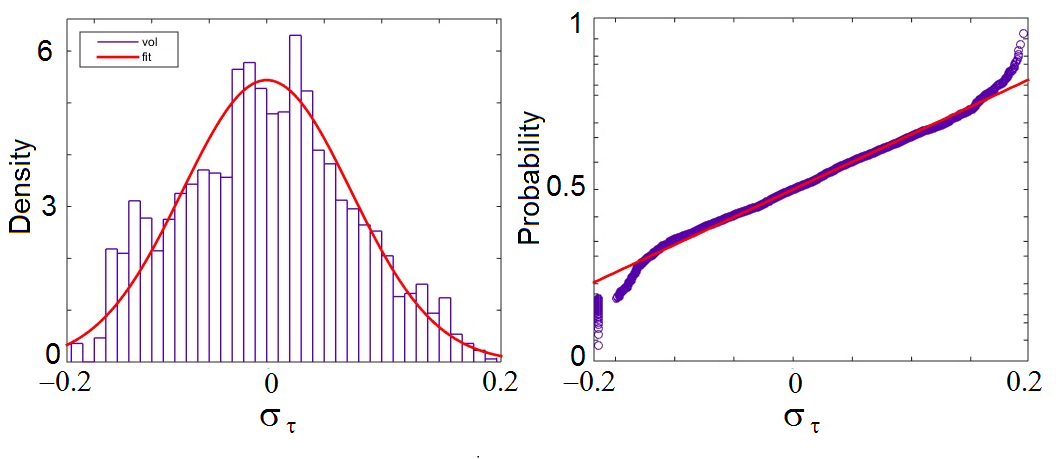}
\caption{Graphical method for comparing the centered $\sigma_\tau$ probability distribution in comparison with a normal distribution (probability density function ($PDF$) and probability plot).} 
\label{fig:fig1}
\end{figure}

From the computed $\kappa$ time series (Eq.\ref{eq:csi}), another series reporting on its intrinsic variability (or volatility $\sigma_{\tau}(t)$) and highlighting the concept of predictive risk is built. To this end, we suggest to use the standard deviation of the $\kappa$ return  ($r(t)=\kappa(t)-\kappa(t-1)$) computed over the $\tau$-sliding windows ($\tau \in \mathbb{N}_{>1}$) (Eq.\ref{eq:eq5}), 
\cyr{
\begin{equation}
\label{eq:eq5}
\sigma_{\tau}(t)=\sqrt{\frac{1}{\tau}\sum_{i=0}^{\tau-1}\bigg(r(t-i)-\frac{1}{\tau}\sum_{n=0}^{\tau-1}r(t-n)\bigg)^2}
\end{equation}}
where $\tau=30$, because for Ajaccio a $30h$ window provides the best results. In the literature, other definitions can be found for the volatility \cite{krawiecki_volatility_2002,kaizoji_statistical_2005}  using in particular the \textit{logarithm} or the \textit{absolute-value} norm. However, here, the given definition yields the best results and constitutes the simplest way to establish the volatility. Fig.\ref{fig:fig1} shows that the trend of the centered $\sigma_{\tau}$ (i.e. volatility minus its mean) distribution for Ajaccio can be considered as normal shape with a slight platykurtic tendency (confirmed with the Jarques-Bera test at the 10\% significance level). 
Rather than working separately on $\kappa$ and $\sigma_{\tau}$, we propose to build $z=\{\kappa(t)+j\sigma_{\tau}(t),t\in \mathbb{Z},j^2=-1 \}$, a scalar complex-valued time series and to model this discrete stochastic process with an autoregressive process of order $p$ ($AR(p)$\cite{box_time_1976}). This model involves random variables defined on the same sample and event spaces and  with the same probability measure (that makes it possible to define distribution function $F$). This method is to be compared to that exposed by Ivan Svetunkov concerning the logic of Brown's exponential smoothing methods and the complex-valued time series used to forecast two-time series simultaneously \cite{svetunkov_complex-valued_2012}, with the difference that there is no volatility issue. From now on, only mean-centered variables will be considered, but will not be introduced in the following equations for readability reasons. 

The complex-valued transform replaces a system of equations related to the prediction of $\kappa$  and its volatility $\sigma_\tau$ (Eq.\ref{eq:decomp1} with $\widehat{(.)}$ for predicted values) by a single regression equation (Eq.\ref{eq:decomp2}, the proof is obvious setting $\boldsymbol{\omega}= \boldsymbol{\xi}+j\boldsymbol{\zeta}$, $\boldsymbol{\omega} \in \mathbb{C}$ and $\boldsymbol{\xi},\boldsymbol{\zeta}\in \mathbb{R}$).    
\begin{subequations}
\begin{equation}
\label{eq:decomp1}
	\left\lbrace
		\begin{aligned}
			\widehat{\kappa}(t+1)=\sum_{i=0}^{p-1}\kappa(t-i)\xi_i-\sum_{i=0}^{p-1}\sigma_{\tau}(t-i)\zeta_i\\
			\widehat{\sigma}_{\tau}(t+1)=\sum_{i=0}^{p-1}\kappa(t-i)\zeta_i+\sum_{i=0}^{p-1}\sigma_{\tau}(t-i)\xi_i
			\end{aligned}
	\right.
\end{equation}
\begin{equation}
\label{eq:decomp2}
\widehat{z}(t+1)=\sum_{i=0}^{p-1}z(t-i)\omega_i
\end{equation}
\end{subequations}

Before using an autoregressive ($AR$) model, it is important to deal with the model identification (the choice of the parameter or order $p$ in Eq.\ref{eq:decomp2}); a classical tool widely studied in regression analysis is employed. It handles with the interpretation of the partial autocorrelation factor ($\beta$ \cite{box_time_1976}) according to real and imaginary parts of $z$ (respectively $\Re (z)$ and $\Im (z)$) as described in, 
\begin{equation}
\label{eq:eq4}
 \exists ~p  \ \vert  \ \beta(t,t-p)\neq 0  \ \textrm{and} \ \beta(t,t-p-1)=0 ~\forall \{t>p\}\in \mathbb{Z} \cite{DEGERINE200346,pal_correlation_2018}
\end{equation}
where $p_\Re$, $p_\Im$ and $p$ denote $AR$ orders having connections with $\Re (z)$, $\Im (z)$ and $z$. We are setting $p=\max(p_\Re,p_\Im)$ to make the problem easier. It is better to benefit from an excess than from a lack of information while referring to the bias-variance trade-off and being aware that the number of inputs should not be too large ($\cyr{p \in [2,6]}$ \cyr{for all the horizons concerning} the studied site). \cyr{By setting this rule, we have neglected the particular algebra imposed by the use of complex numbers, but there is, to our knowledge, no other way of doing it that would be as simple.} Note that dealing with deseasonalized series $\kappa$, $p$ does not have to be greater than $24h$ to capture the daily seasonal cycle. There are other identification methods, as for example the complex autocorrelation factor \cite{gubner_probability_2006}, however, it seems that this method has a worse ratio complexity-efficiency. 
The next step is the model estimation ($\boldsymbol{\omega}$) by transposing what has been done for many years in the real-valued case (least square optimization) to the complex-valued case. Considering an input matrix $\mathbf{I}\in \mathbb{C}^{D \times p}$ (Eq.\ref{eq:I}) and an output column vector $\mathbf{o}\in \mathbb{C}^{D \times 1}$ (Eq.\ref{eq:o}), the solution of the $AR(p)$ least squares problem consists in determining unknown parameters ($\boldsymbol{\omega}\in \mathbb{C}^{p \times 1}$ in Eq.\ref{eq:o}). By the way, the problem, already raised and well detailed in \cyr{the paper of Adrian et al.} \cite{adrian_complex-valued_2018} for spatial data-based model is resumed as in the classical real-valued case by $\mathbf{I}\boldsymbol{\omega}=\mathbf{o}$ and can be solved from the formulation of the mean square error estimation ${\mathbb{E}[\mathbf{e}}^2]=\left\|\mathbf{I}\boldsymbol{\omega}-\mathbf{o}\right\|^2$ \cite{PhysRevA.94.022342}. 
Note that the authors of this paper do not venture to state that the least squares method provides the best solution to the problem.
To be able to assert it, one has to prove that there is equivalence with the maximum likelihood and have to formulate hypotheses on the complex residual distribution.
\begin{align}
\mathbf{I} & = 
\left(
\begin{array}{cccc}
{z}(t-1) & {z}(t-2) & \cdots & {z}(t-p)\\
{z}(t-2) & {z}(t-3) & \cdots & {z}(t-p-1)\\
\vdots & \vdots &   & \vdots \\
{z}(t-D) & {z}(t-D-1) & \cdots & {z}(t-D-p+1)\\
\label{eq:I}
\end{array}
\right)_{D\times p}\\
\boldsymbol{\omega} & = \big(\omega_{1},\ldots ,\omega_{p}\big)_{p\times 1}^{\prime}, \mathbf{o}=\big(z(t),\ldots ,z(t-D+1)\big)_{D\times 1}^{\prime}
\label{eq:o}
\end{align} 

\begin{figure}
\includegraphics[scale=0.30]{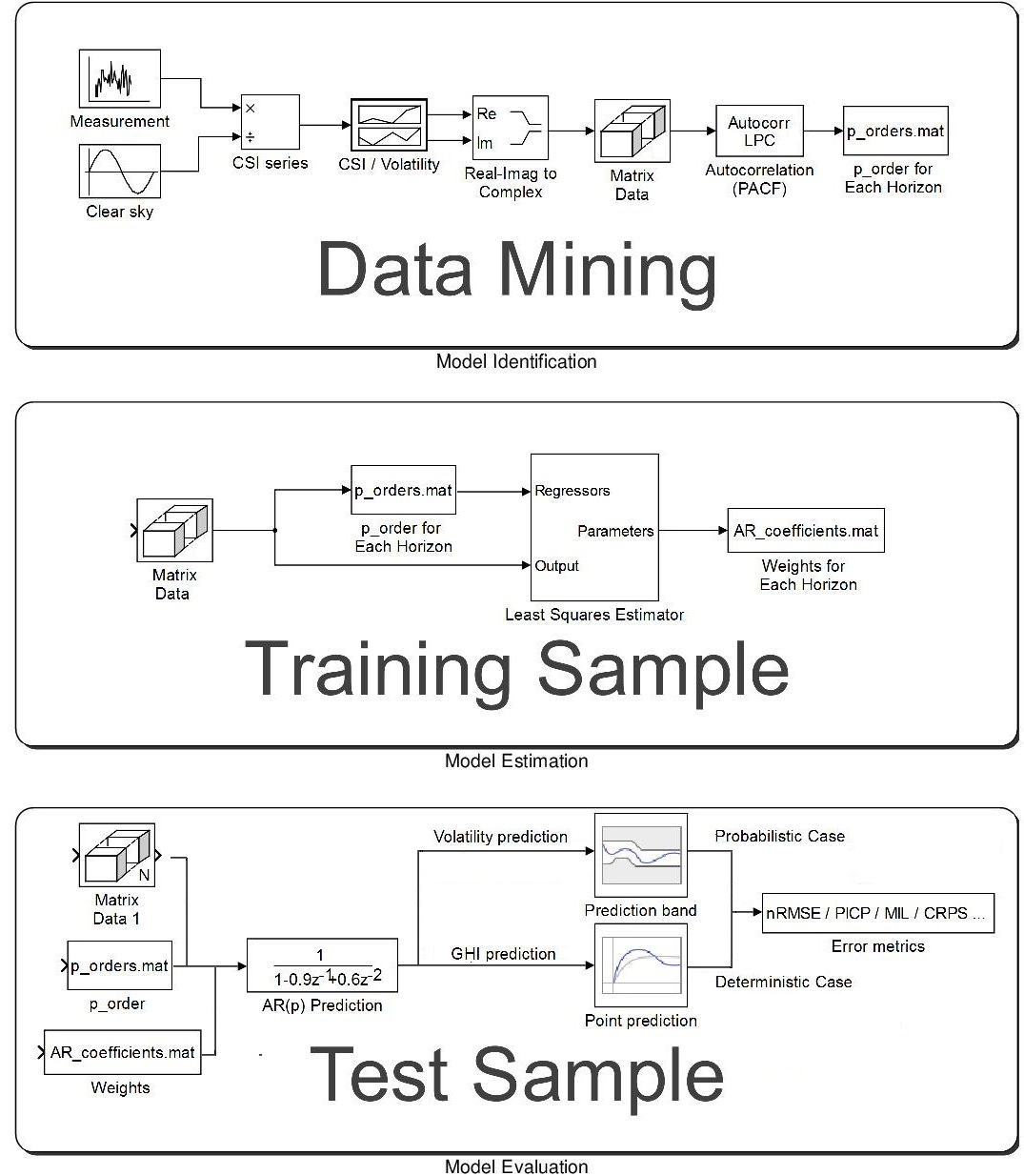}
\caption{Conceptual diagram of the experiment.}
\label{fig:fig2}
\end{figure}

The complex-valued case differs from the real-valued one, replacing the \cyr{$L_2$-norm} by the Frobenius norm introducing the Frobenius inner product \cite{PhysRevE.91.012820} on $\mathbb{C}^D$   (${\mathbb{E}[\mathbf{e}}^2]=<(\mathbf{I}\boldsymbol{\omega}-\mathbf{o}),(\mathbf{I}\omega-\mathbf{o})>_F$). Classically, the minimum of the squared expected value $({\mathrm{argmin}} ({\mathbb{E}[\mathbf{e}}^2]):=\{\boldsymbol{\omega}\in\mathbb{C}^p | \forall \boldsymbol{\omega}^*\in\mathbb{C}^p: {\mathbb{E}[\mathbf{e}}^2(\boldsymbol{\omega}^*)]\geq{\mathbb{E}[\mathbf{e}}^2(\boldsymbol{\omega})])$ is carried out computing its differentiating (Eq.\ref{eq:eq1}) and by letting $\partial {\mathbb{E}[\mathbf{e}^2]}/\partial( \boldsymbol{\omega}^H)=0$ where $(.)^H$ defines conjugate transpose.
\begin{equation}
\label{eq:eq1}
		\begin{aligned}
			\frac{\partial{\mathbb{E}[\mathbf{e}}^2]}{\partial(\boldsymbol{\omega}^H)}&=\frac{\partial(I\boldsymbol{\omega}-\mathbf{o})^H(I\boldsymbol{\omega}-\mathbf{o})}{\partial(\boldsymbol{\omega}^H)}\\
			&=	\frac{\partial(\boldsymbol{\omega}^HI^H-\mathbf{o}^H)(I\boldsymbol{\omega}-\mathbf{o})}{\partial(\boldsymbol{\omega}^H)}
			\end{aligned}
\end{equation}
It must be emphasised that $\partial {\mathbb{E}[\mathbf{e}}^2]/\partial (\boldsymbol{\omega}^H)$ is the complex conjugate transpose of $\partial {\mathbb{E}[\mathbf{e}}^2]/\partial \boldsymbol{\omega}$, thus, setting one to zero also sets the other to zero. The normal equation, in this complex-valued case, becomes,
\begin{equation}
\label{eq:eq2}
\mathbf{I}^H(\mathbf{I}\boldsymbol{\omega}-\mathbf{o})=0
\end{equation}
Furthermore, the solution of this matrix equation corresponds to a regression coefficients \cite{claerbout_geophysical_2014} estimated by $\tilde{\boldsymbol{\omega}}=(\mathbf{I}^H\mathbf{I})^{-1}\mathbf{I}^H\mathbf{o}$ . Therefore, differentiating by a complex-valued vector is an abstract concept,  but  it  yields  the  same  set  of  equations as differentiating separately each scalar component (real and imaginary) and is a more concise form \cite{van_den_bos_estimation_1994}. 
On top of that, to improve the condition number of the problem, one can introduce a constrained minimization with $\left\|\boldsymbol{\omega}\right\|^2<r(\lambda)$ where $r$ is a bijective function and $\lambda$ is the Lagrange multiplier of the constraint ($\mathbf{I}^H\mathbf{I}+\lambda^H\lambda$) defined positive and so invertible. This approach denoted \texttt{Ridge} approach \cite{ridge} can also be used in the complex-value case, in the form $\tilde{\boldsymbol{\omega}}_{\lambda}=(\mathbf{I}^H\mathbf{I}+\lambda \mathds{1})^{-1}\mathbf{I}^H\mathbf{o}$. A machine learning-like approach consists in performing cross-validation and selecting the $\lambda$ value that minimizes the out-sample sum of squared residuals; in our experimental setup, $\lambda=3.74$ is the best choice (i.e. inducing the lowest prediction errors). Now the identification and optimization problems have been analyzed (the outcome of the experiment is detailed in Fig.\ref{fig:fig2}), it is required to theoretically validate the use of predictions of both the $GHI$ and its volatility in the case of the probabilistic forecasting \cite{gneiting_probabilistic_2007,app12041823}.
 The goal is to show that the volatility detailed previously could be used to capture the idea of unpredictable and quick fluctuations. 
 Thereby, considering the $\ell$-step head prediction, it would be possible given a well-chosen $\mu_{t+\ell}$  parameter, to bound the prediction considering that $GHI(t+\ell)$ measurement is included in the interval $\Lambda$ satisfying the Eq.\ref{eq:fliess}. The following is dedicated to answering the question: can we theoretically explain $\mu_{t+\ell}$?  
\begin{equation}
\label{eq:fliess}
\begin{aligned}
	\Lambda=[\widehat{GHI}(t+\ell)- \mu_{t+\ell}GHI_{CS}(t+\ell) \widehat{\sigma}_\tau(t+\ell),\\ \widehat{GHI}(t+\ell)+ \mu_{t+\ell} GHI_{CS}(t+\ell)\widehat{\sigma}_\tau(t+\ell)]
\end{aligned}
\end{equation}
  
  First of all, it is important to consider this study within non-standard analysis framework with $S$-integrable time series (additive decomposition \cite{lobry_non-standard_2008}) and by referring to the Cartier-Perrin theorem \cite{cartier_integration_1995}, as suggested in many papers from Fliess \cite{fliess_volatility_2011,fliess_prediction_2018}. In this context, we can explore the fact that our prediction method (and more generally all the machine learning approaches) only predicts the trend of the $GHI$ but certainly not fast fluctuations. From Eq.\ref{eq:eq5}, it is conceivable to interpret the volatility thanks to the Backshift operator $\boldsymbol{B}$; it holds,  
\begin{equation}
\label{eq:sigma}
		\begin{aligned}
			\sigma_{\tau}^2(t)&=\mathbb{E}[r^2]-\mathbb{E}[r]^2\simeq\mathbb{E}[r^2]=\mathbb{E}[(\kappa-\boldsymbol{B}^1\kappa)^2]\\
			               &=\frac{1}{\tau}\sum_{i=0}^{\tau-1}\bigg(\kappa(t-i)-\kappa(t-i-1)\bigg)^2
			\end{aligned}
\end{equation}
where over a sufficiently large interval, $\kappa(t)$ oscillates around a constant mean value, making the average of the return close to $0$ ($\mathbb{E}[r]^2 \underset{\tau\to +\infty}{\longrightarrow} 0$) and $\boldsymbol{B}^\ell \kappa(t)= \kappa(t-\ell)$.  
This equation is not unlike the classical formulation of the variance in which the mean of $\kappa$ is replaced by the $\kappa(t-i-1)$. Of course, it sounds appealing to propose probabilistic prediction using the variance of residual; but in $GHI$ prediction, the Gaussian hypothesis is never verified \cite{trapero_calculation_2016,voyant_periodic_2018}. It is also known that the prediction intervals are too wide and become quickly unusable with the hypothesis of the persistence of the variance \cite{fliess_volatility_2011}. Furthermore, the option of proposing increasingly complex non-parametric methods is satisfactory from a theoretical point of view but it is not very advantageous in practice. There are a lot of interpretations of Eq.\ref{eq:sigma} and the attentive reader will recognize the formulation of the mean square error with respect to a persistence model or the fact that $\sigma_{\tau}^2(t)=\mathbb{E}[\big(\frac{\partial \kappa}{\partial t}\big)^2]$. With this equation, it is tempting to believe that we have stumbled upon a deep difficulty, but a $\kappa$ breakdown into a trend ($T$) and fast fluctuations terms ($\epsilon$) transforms Eq.\ref{eq:sigma} into,
\begin{equation}
\label{eq:sigma2}
		\begin{aligned}
			\sigma_{\tau}^2(t)=\frac{1}{\tau}\sum_{i=0}^{\tau-1}&\bigg((T(t-i)-\mathbb{E}[T])-(\boldsymbol{B}^1T(t-i)-\mathbb{E}[T])\\
			&+(\epsilon(t-i)-\mathbb{E}[\epsilon])-(\boldsymbol{B}^1\epsilon(t-i)-\mathbb{E}[\epsilon])\bigg)^2
			\end{aligned}
\end{equation}
considering that $\kappa(t)=T(t)+\epsilon(t)$.
To go further, we shall consider the co-variance $\sigma_\epsilon(t,t-1)$ and the partial  autocorrelation functions which is identical to the autocorrelation function for the lag $1$ $\beta_\epsilon(t,t-1)$ for dependency between $\epsilon$ and himself $1$ lag delayed. By contrast with the standard analysis, here the high-frequency term ($\epsilon$) has no mean and co-variance functions tending to 0 ($\beta_T$ is a function close to 1 and $\sigma^2_T<<\sigma^2_{\epsilon}$ ) which means that Eq.\ref{eq:sigma2} can be replaced by,
\begin{equation}
\label{eq:var}
		\begin{aligned}
			\sigma_{\tau}^2(t)&=2\sigma^2_\epsilon(t)-2\sigma_\epsilon(t,t-1)+2\sigma^2_T(t)-2\sigma_T(t,t-1)\\
                   &=2\sigma^2_\epsilon(t)\big(1-\beta_\epsilon(t,t-1)\big)+2\sigma^2_T(t)\big(1-\beta_T(t,t-1)\big)\\
		\end{aligned}
\end{equation}

Bearing in mind the above, this equation shows that there is a link between the volatility as described in Eq.\ref{eq:eq5} and the variance of the high-frequency component ($\epsilon$). By means of the König-Huygens' theorem and characteristics of linear correlation coefficient of Bravais-Pearson with the fact that the covariances between $T$ and $\epsilon$ are close to 0 (considering $T\perp \epsilon$ with $\mathbb{E}[\epsilon]=0$), we may show that Eq.\ref{eq:var} could be replaced by the more practical expression $\sigma_{\tau}^2(t)\simeq 2\sigma_{\epsilon}^2(t)\big(1-\beta_{\epsilon}(t,t-1)\big)$. It is possible to settle $T$ with a classical  moving average defined by a (2n+1)-point mean values: $T(t)=\mathbb{E}[\kappa(t-n:t+n)]$ and $\epsilon$ with $\epsilon(t)=\kappa(t)-T(t)$. With the daytime filtering process (Section \ref{sec:data}) $n=5$ provides a daily average. It will be needful to introduce the transform $\Gamma:\sigma_{\tau}\in [0,1]\to \Gamma(\sigma_{\tau})\in \mathbb{R}^+$ in order to handle with $\sigma_{\epsilon}^2$ (Eq.\ref{eq:cool} with $\beta < 1$, $\sqrt(2)\sigma_{\epsilon}=\Gamma(\sigma_{\tau})$).
\begin{equation}
\label{eq:cool}
			\Gamma(\sigma_{\tau})=\frac{\sigma_{\tau}}{\sqrt{1-\beta_{\epsilon}(t,t-1)}}
\end{equation}

Once the prediction $\widehat{z}(t+1)$ is obtained, it is easy to compute next value of the $\widehat{\kappa}(t+1) = \Re\left[\widehat{z}(t+1)\right]$ and the associated volatility  $\widehat{\sigma}_{\tau}(t+1)=\Im\left[\widehat{z}(t+1)\right]$.
From here, we propose to build an estimate of the probabilistic $GHI$ prediction based on the point prediction ($\widehat{GHI}(t+1)=\widehat{\kappa}(t+1)GHI_{CS}(t+1)$) and the cumulative distribution function ($F_{\kappa}$) computed from the conditional volatility: $F_{\kappa}(x)=\mathbb{P}(\kappa<x)$ \cite{deisenroth_mathematics_2020}. This last term corresponds to the probability that the random $\kappa$ variable takes on a value less than or equal to $x$. The probability that $\kappa$ lies in the semi-closed interval $( a , b ]$, is therefore $\mathbb{P}(a<\kappa \leq b)=F_{\kappa}(b)-F_{\kappa}(a)$. In the Gaussian case assumed here (the quantities \textit{mean}, \textit{expectation}, \textit{median} and \textit{mode} of the distribution are identical \cite{krishnamoorthy_handbook_2006}), $F_{\kappa}$ and his inverse $F^{-1}_{\kappa}$ are defined from the error function ($\erf$) as described respectively in, 
\begin{equation} 
\label{eq:F}
\widehat{F}_{\kappa}(x)=\frac{1}{2}+\frac{1}{2}\erf\bigg(\frac{x-\widehat{\kappa}}{\Gamma(\widehat{\sigma}_{\tau})}\bigg) 
\end{equation}
where $x\in \mathbb{R}$ and,
\begin{equation}
\label{eq:F-1}
\widehat{F}^{-1}_{\kappa}(q)=\widehat{\kappa}+\Gamma(\widehat{\sigma}_{\tau})\erf^{-1}(2q-1)
\end{equation}
where $0<q<1$. Probabilistic forecasting is more powerful than the deterministic one and allows us to bound the prediction proposing that is called prediction interval from quantiles estimation at probability level $q\in[0,1]$ $\widehat{Q}(q)=inf \{x\in \mathbb{R}: \widehat{F}_{\kappa}(x)\geq q\}$. Consider that, if the the function $\widehat{F}$ is continuous and strictly monotonically increasing, we have the quantile function defined by {$\widehat{Q}(q)=\widehat{F}^{-1}(q)$} (denoted \texttt{probit} function in the Gaussian case)\cite{hyndman_sample_1996}. In that instance of a central prediction interval (the most common way is to center the prediction interval on the median considering there is the same probability of risk below and above the median \cite{pinson_2007}). with a nominal coverage rate of $(1-\alpha)100\%$, the lower bound ($\underline{GHI}$) is estimated by using the $\alpha/2$ quantile and the upper bound ($\overline{GHI}$) using the $1-\alpha /2$ quantile as described in Eq.\ref{eq:bound} with an example quantile function estimation in the normal distribution case ($\erf^{-1}$ is an odd function).
\begin{equation}
\label{eq:bound}
	\left\lbrace
		\begin{aligned}
			\underline{GHI}&=\widehat{Q}(\alpha/2)=\widehat{GHI}-\erf^{-1}(1-\alpha)GHI_{CS}\Gamma(\widehat{\sigma}_{\tau})\\
			\overline{GHI}&=\widehat{Q}(1-\alpha/2)=\widehat{GHI}+\erf^{-1}(1-\alpha)GHI_{CS}\Gamma(\widehat{\sigma}_{\tau})
			\end{aligned}
	\right.
\end{equation}

Point out that as this is very frequently done in solar irradiance prediction, these interval limits can in turn be limited by considering that the upper limit is necessarily lower than $GHI_{CS}$ and that the lower limit shall be higher than the diffuse component of the $GHI_{CS}$ (this quantity is easily obtained with the \texttt{Solis} modeling) \cite{fliess_prediction_2018}. We previously treated the $t+1$ case, nevertheless the reasoning for $\mu_{t+\ell}$ is rather similar replacing (in Eq.\ref{eq:decomp2}) $\widehat{z}(t+1)$ by $\widehat{z}(t+\ell)$. Assuming all the approximations made so far (normal assumption of $\sigma_{\tau}$ in Eqs.\ref{eq:F} and \ref{eq:F-1}, the arbitrary choice of $\tau$ in Eq.\ref{eq:eq5} and the hypothesis on $\mathbb{E}[r]^2$ in Eq.\ref{eq:sigma}), it is doable and advisable to calibrate the $\mu_{t+\ell}$ value in Eq.\ref{eq:fliess} according to nominal coverage rate $(1-\alpha)100\%$ by performing simulations on the training space (link between $(1-\alpha)100\%$ and $\mu_{t+\ell}$ values). Furthermore, in Table I, it is shown that the two approaches lead to quite different results and that Eq.\ref{eq:bound} shall only be considered as a first approximation requiring data-driven corrections (see Annex \ref{annex} for details). 

\begin{table}[b]
\label{table:table1}
\caption{$\alpha_{t+1}$ estimations from Eq.\ref{eq:bound} and data guided approach performing simulations on the training space (data driven correction for $1h$ horizon and a 11-point mean values corresponding to $n=5$ and  $\beta_{\epsilon}(t,t-1)=0.38$).
}
\begin{ruledtabular}
\begin{tabular}{ccccc}
&$\bm{\alpha=0.2}$&$\bm{\alpha=0.4}$&$\bm{\alpha=0.6}$&$\bm{\alpha=0.8}$\\
$\alpha_{t+1}$ (Eq.\ref{eq:bound})\footnote{coupling Eqs\ref{eq:cool} and \ref{eq:bound} $\alpha_{t+1}$ can be found like equal to $(1-\beta_{\epsilon}(t,t-1))^{-1/2}\erf^{-1}(1-\alpha)$ }  &1.15 &0.76  &0.47  &0.23  \\
Data driven\footnote{these data can be fitted with an exponential decay ($R^2=0.999$) according to $\alpha_{t+1}=1.916e^{-3.034\alpha}$
}  &1.04  &0.57  &0.31  &0.17  
\end{tabular}
\end{ruledtabular}
\end{table}

\section{Results}
\label{sec:results}
Despite the fact that the purpose of this paper is to elaborate a new way to propose $GHI$ probabilistic forecast (complex-valued method denoted $Compl$), it is important to compare results with some classical tools, like a Gaussian parametric process (denoted $Gauss$ and based on the variance of the residual \cite{voyant_periodic_2018}), a non-parametric bootstrapped process (denoted $Boot$ \cite{pan_bootstrap_2016}) and a Ridge quantile regression model (denoted $Quant$ \cite{lauret_probabilistic_2017,van_der_meer_review_2018,CARNEIRO2022118936}). The used error metrics for the comparison in the deterministic case is  normalized root mean square error ($nRMSE \ $\cite{VOYANT2022747}) while in the probabilistic case, we choose normalized mean interval length ($MIL$ sometime denoted $PINAW$ for prediction interval normalized average width), percentage interval coverage probability ($PICP$), continuous rank probability score ($CRPS$) and mean scaled interval score ($MSIS$). All these metrics are described in \citet{lauret_verification_2019,hyndman_another_2006,van_der_meer_review_2018} and references therein. In Table II is shown the comparison between all the prediction interval methodologies and is proved that the complex approach is equivalent in terms of deterministic prediction ($nRMSE$ nearly identical for all five methods) but grants, considering a nominal coverage rate of 80\%, a significant $MIL$ decrease that is worthwhile for a grid operator who seeks a predictive methodology offers the lowest conceivable $MIL$ for a given $PICP$. Another interesting element is the fact that for both $Quant$ and $Compl$ methods, $\alpha=0.2$ (nominal covering rate of $100\%(1-0.2)=80\%$) effectively corresponds to a $PICP$ close to $80\%$ unlike the two other cases.  

\begin{table}[b]
\label{tab:table2}
\caption{
Models comparison for a nominal coverage probability of 80\% ($\alpha=0.2$)
}
\begin{ruledtabular}
\begin{tabular}{clcccc}
\textbf{Horizons}&\textbf{Metrics}&\textbf{Gauss}&\textbf{Boot}&\textbf{Quant}&\textbf{Compl}\\
1h &\textit{nRMSE}  &{0.197} &{0.203}  &{0.201}  &{0.196}  \\
                  &\textit{PICP(\%)}  &83.81  &75.21  & 79.65 &80.01  \\
                  &\textit{MIL(\%)}  &51.24  &40.36  & 42.24 &41.24  \\
2h &\textit{nRMSE}  &{0.251} &{0.267}  &{0.258}  &{0.252}  \\
                  &\textit{PICP(\%)}  &81.39  &76.67  & 79.90 &80.74  \\
                  &\textit{MIL(\%)}  &64.57  &55.47  & 58.72 &56.72  \\
3h &\textit{nRMSE} &{0.282}  &{0.304}  &{0.289}  &{0.282}  \\
                  &\textit{PICP(\%)}  &80.68  &73.27  & 80.07 &80.06  \\
                  &\textit{MIL(\%)}&70.97  &58.22  & 67.32 &63.09  \\
4h &\textit{nRMSE} &{0.302}  &{0.327}  &{0.312}  &{0.303}  \\
                  &\textit{PICP(\%)}  &80.42  &81.69  & 80.64 &80.02  \\
                  &\textit{MIL(\%)}&75.39  &76.82  & 74.47 &67.10  \\
5h &\textit{nRMSE}  &{0.316}  &{0.362}  &{0.328} &{0.317}  \\
                  &\textit{PICP(\%)}  &80.99  &74.44  &80.81  &79.49  \\
                  &\textit{MIL(\%)}&78.73  &64.45  & 79.08 &69.32  \\
6h &\textit{nRMSE} &{0.324}  &{0.358}  &{0.339}  &{0.325}  \\
                  &\textit{PICP(\%)}  &81.40  &78.69  &81.43  &79.66  \\
                  &\textit{MIL(\%)}&81.41  &73.79  & 82.73  &70.05 
\end{tabular}
\end{ruledtabular}
\end{table}

In Fig.3, one can observe how $Compl$ forecast intervals are distributed considering $1h$ horizon. 
\begin{figure*}
\label{fig:fig3}
\includegraphics[scale=0.46]{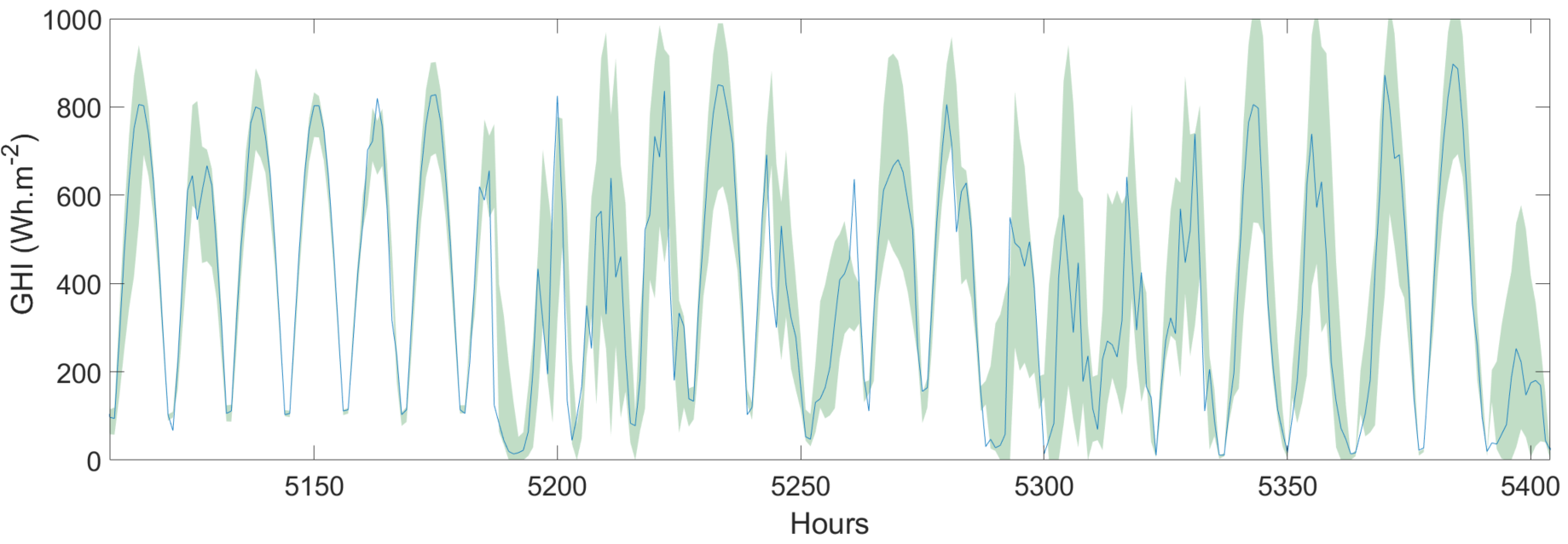}%
\caption{80\% prediction interval with complex-valued approach versus measures (blue line) }
\end{figure*}
The main attraction of the method lies in the fact that the interval band is conditioned by the variability observed in the previous hours. Thus, for the days close to the $5150^{th}$ the prediction band (very small) is completely different from what is observed close to the $5350^{th}$ hours (very large). The probabilistic counterpart of the mean absolute error is the $CRPS$, making it possible to quantify the total error made with the predicted distributions as it is shown in Fig.4. 
\begin{figure}
\label{fig:fig4}
\includegraphics[scale=0.25]{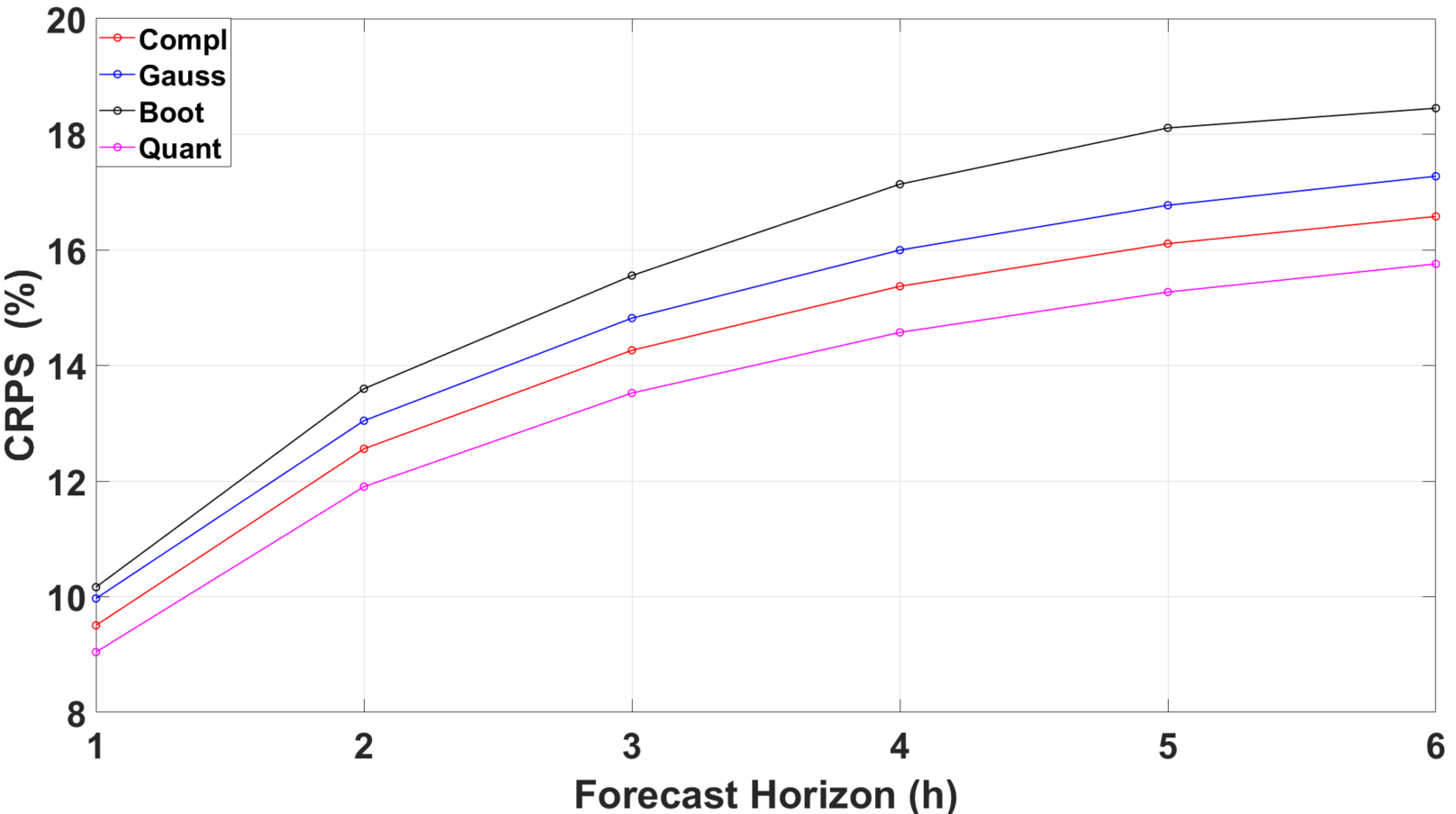}%
\caption{$CRPS$ for the probabilistic comparison}
\end{figure}
Thence, it is a robust score that is designed in such a way that it measures both reliability and sharpness. An advantage of the $CRPS$ is that it reduces the absolute error if the forecast is deterministic, and allows the comparison between probabilistic and point forecasts \cite{van_der_meer_review_2018}. We may note that even if the quantile regression is the best tool considering this metric, the errors observed by the complex-valued methodology are not prohibitive. This phenomenon is also visible by comparing the $ MSIS $ (related to $\alpha=0.2$) which has the enormous advantage of considering all the forecast horizons within a single metric. If for $Boot$ and $Gauss$, $MSIS$ are respectively 1.05 and 1.03, for $Quant$ and $Compl$, $MSIS$ are lower and so better (0.89 and 0.95).

\section{Conclusions}
\label{sec:Conclusions}

The objective of this paper is to present a new method for predicting $GHI$ that is able to take into account fast fluctuations. Often the literature boasts some sophisticated approaches, but when focusing on the existing installations, one remarks that the highly-developed models yield way to simpler methods. Although less effective, they are more robust and easier to use. From a practical point of view, a ``good" method concerns a tool that would be easily usable in a stand-alone application (problems of some toolboxes), and which doesn't involve a lot of different concepts or data. The procedures used for smart management shall be self-sufficient and consistent with continuous learning and with some eventual detectors failure. It is in this perspective we tested a new univariate methodology based on the complex-valued time series generated from $GHI$ measurements. With only a few parameters ($6$ complex numbers in the studied case) and some basic mathematical operations, this approach makes it possible to predict $GHI$ with accuracy compared with classical probabilistic and deterministic predictions. This method proposes the lowest $MIL$ considering a fixed nominal coverage rate ($80\%$). Once the parameters have been estimated and provided that real-time $GHI$ measurements are available, a simple spreadsheet can become a tool of choice in the management of $PV$ installations. 
The validation of this approach will require many more tests by varying time steps, horizons, and forecastability \cite{doi:10.1063/5.0042710} or predictability \cite{doi:10.1063/5.0056918}. However, this new forecast methodology is simple to implement and may facilitate the integration of renewable energies and improve the management of installations using solar radiation as energy sources (smart grid, building, district, etc.). Interesting perspectives will be to apply it to other kinds of time series (not necessarily in connection with renewable energies), to construct the imaginary part concerning other variables than volatility (residuals, exogenous or ordinal data, etc.), and perhaps adapt the method to others predictors kinds (artificial neural network, support vector regression, etc.).    

\parskip=10pt
\textbf{Conflict of Interest.}
The authors have no conflicts to disclose.

\parskip=10pt
\textbf{Data Availability.}
The data that support the findings of this study are available
from the corresponding author upon reasonable request.

\appendix*
\section{Data Driven Correction}
\label{annex}
The data-driven method proposed here allows for improving probabilistic forecasting. From Eq.\ref{eq:fliess}, it would be useful to determine experimentally (and not theoretically with Eq:\ref{eq:bound}) $\mu_{t+\ell}$ such as $\mathbb{P}(GHI\in\Lambda)\to (1-\alpha)$ when the number of observations is large enough. All along the training step, curves fitting to inverse cumulative distribution functions are fixed ($\mu_{t+\ell}$ as a function of $\alpha$). Its use requires a few assumptions (less than in the theoretical case presented above in this paper). The fine advantage lies in the fact that the Gaussian hypothesis no longer has any reason to exist (non-parametric method). Nonetheless, two new much less restrictive hypotheses must be formulated. The first one is that there is the same probability of the risk below and above the median (common postulation \cite{pinson_2007}) and the second one is that the $\sigma_\tau$ distribution is symmetric (the mean and the median are identical). The sample skewness is worth $0.1$, hence it is regular to consider the second assumption as verified (since comprised between $-1$ and $1$ \cite{george_2003}). In Table III are shown the $f_1$ and $f_2$ parameters values concerning the fit $\mu_{t+\ell}=f_1e^{f_2\alpha}$.
\begin{table}[b]
\label{tab:table3}
\caption{
$\mu_{t+\ell}$ adjustment for each horizon $\ell$ ($f_1$ and $f_2$ the constants of the exponential decay fit)
}
\begin{ruledtabular}
\begin{tabular}{cccc}
$\bm{\ell}$&$\bm{f_1(CB95\%)}$\footnote{estimates with 95$\%$ confidence bounds}&$\bm{f_2(CB95\%)^\textrm{a}}$&$\bm{R^2}$\footnote{coefficient of determination}\\
$1$& 1.916(1.745,2.087)& -3.034(-3.322,-2.747)  &0.999  \\
$2$& 2.739(2.705,2.773)&-3.163(-3.218,-3.109) &0.994    \\
$3$& 2.828(2.797,2.860)& -2.811(-2.853,-2.769) &0.995   \\
$4$& 2.869(2.834,2.904)& -2.605(-2.647,-2.563) &0.994   \\
$5$& 2.821(2.787,2.855)& -2.393(-2.432,-2.354) & 0.993    \\
$6$& 2.707(2.673,2.741)& -2.175(-2.214,-2.136) & 0.992    \\
\end{tabular}
\end{ruledtabular}
\end{table}
This data-driven method may be used to estimate the quantiles ($\widehat{Q}$) and so the cumulative distribution function. Indeed, considering $\Delta q\in [0,0.5]$, we assume,
\begin{equation}
\label{eq:fit}
	\left\lbrace
		\begin{aligned}
			\widehat{Q}(0.5+\Delta q)&=\widehat{Q}(0.5)+f_1e^{f_2(1-2\Delta q)}\widehat{\sigma}_{\tau}\\
			\hat{Q}(0.5-\Delta q)&=\widehat{Q}(0.5)-f_1e^{f_2(1-2\Delta q)}\widehat{\sigma}_{\tau}\\
			\widehat{Q}(0.5)&=\widehat{GHI}
			\end{aligned}
	\right.
\end{equation}
verified if and only if, $\widehat{Q}$ is a continuous function, which implies Eq.\ref{eq:continuity} and thereby $f_1e^{f_2}\to 0$.  
\begin{equation}
\label{eq:continuity}
		\lim\limits_{\substack{\Delta q \to 0^+ \\\Delta q \to 0^-}}\widehat{Q}(0.5\pm\Delta q)=\widehat{Q}(0.5)
		\end{equation}

Taking a concrete example, quantiles $\widehat{Q}(0.1)$ and $\widehat{Q}(0.9)$ could be respectively estimated from a nominal $80\%$ prediction interval ($\alpha=0.2$ and $\Delta q=0.4$) with $\widehat{Q}(0.5)-f_1e^{f_20.2}\sigma_{\tau}$ and $\widehat{Q}(0.5)+f_1e^{f_20.2}\sigma_{\tau}$. To slightly improve the results, and to position oneself in a totally non-parametric approach, it is doable to use lookup tables rather than curve fitting.

\parskip=10pt
\textbf{ACKNOWLEDGMENTS}

This work was partially supported by ANR grant SAPHIR project ANR-21-CE04-0014-03.

\parskip=10pt
\textbf{REFERENCES}
\bibliography{bib}

\end{document}